\documentclass[paper,floatfix]{revtex4-1}

\usepackage{amsfonts,amssymb,amsmath,array}
\usepackage{graphicx}

\begin{document}

\title{Stochastic Model for a Piezoelectric Energy Harvester Driven by Broadband Vibrations}

\author{Angelo  Sanfelice}
\affiliation{Department of Mathematics and Physics, University of Campania ``Luigi Vanvitelli'',  viale Lincoln 5, 81100 Caserta, Italy}

\author{Luigi Costanzo}
\affiliation{Department of Engineering, University of Campania ``Luigi Vanvitelli'', via Roma 29, 81031 Aversa, Italy}

\author{Alessandro Lo Schiavo}
\affiliation{Department of Engineering, University of Campania ``Luigi Vanvitelli'', via Roma 29, 81031 Aversa, Italy}

\author{Alessandro Sarracino}
\affiliation{Department of Engineering, University of Campania ``Luigi Vanvitelli'', via Roma 29, 81031 Aversa, Italy}

\author{Massimo Vitelli}
\affiliation{Department of Engineering, University of Campania ``Luigi Vanvitelli'', via Roma 29, 81031 Aversa, Italy}

\begin{abstract}We present an experimental and numerical study of a
  piezoelectric energy harvester driven by broadband
  vibrations. {This device can extract power from random
    fluctuations and can be described by} a stochastic model, based on
  an underdamped Langevin equation with white noise, {
    which mimics the dynamics of the piezoelectric material. A crucial
    point in the modelisation is represented by the appropriate
    description of the coupled load circuit that is necessary to
    harvest electrical energy.}  We consider a linear load
  (resistance) and a nonlinear load (diode bridge rectifier connected
  to the parallel of a capacitance and a load resistance), and focus
  on the characteristic curve of the extracted power as a function of
  the load resistance, {in order to estimate the optimal
    values of the parameters that maximise the collected energy.} In
  both cases, we find good agreement between the numerical simulations
  {of the theoretical model} and the results obtained in
  experiments. {In particular, we observe} a non-monotonic
  behaviour of the characteristic curve which signals the presence of
  an optimal value for the load resistance at which the extracted
  power is maximised.
  {We also address a more theoretical
    issue, related to the inference of the non-equilibrium features of
    the system from data: we show that the analysis of} high-order
  correlation functions {of the relevant variables}, when
  in the presence of nonlinearities, can represent a simple and
  effective tool to check the irreversible dynamics.
\end{abstract}

\maketitle

\section{Introduction 
}

{ Energy harvesting is the process of capturing and
  storing energy from various environmental sources and represents an
  important topic for both experimental and theoretical
  studies~\cite{gammaitoni2012there}. From the technological
  perspective,} this field has seen significant advancements in recent
years,
{leading to the realisation of several devices}, such as
photovoltaic cells, wind turbines, piezoelectric devices,
thermoelectric generators, and electromagnetic energy harvesters, designed
to collect energy from the environment~\cite{singh2021energy,liu2022biomechanical,pan2021kinetic}.

{A central issue in this framework is certainly
  represented by the maximisation of the extracted power. Indeed,
  depending on the specific design and on the parameters of the
  considered device, energy sources can be exploited in more efficient
  ways. Such a tuning of the relevant parameters can be worked out
  more easily if a theoretical model can be developed that can
  describe the real system.  Since} in many cases energy harvesters
work under random and uncontrolled conditions, due to the
unpredictable nature of the environmental energy sources, stochastic
processes can play a significant role in the modelling of these systems.  In particular, analytical and numerical studies
of simplified models can be useful to design more robust energy
harvesters that can adapt to the inherent variability of their energy
sources, maximising energy capture.  {Let us also mention that}, from a more general and theoretical
perspective, the problem of rectifying random fluctuations is studied
within the theory of Brownian (or molecular) motors~\cite{R02},
also known as ratchet models, where the presence of a spatial
asymmetry coupled with non-equilibrium conditions allows one to extract
directed motion from unbiased fluctuations~\cite{puglisi13,Leonardo9541}.

Among the several mentioned physical mechanisms exploited to harvest
energy from the environment, we focus here on piezoelectric
materials~\cite{clementi2022review}. These have the properties to
convert mechanical stress into electrical energy, making them good
candidates for vibration energy harvesting applications.{
  Indeed, the electrical currents generated when the material is
  deformed can be collected by suitable load circuits and electrical
  power can be obtained. Devices based on this mechanism are usually
  used to feed small sensors, for instance in wireless sensor
  networks.} {These kinds of energy harvesters are
  generally constituted by a piece of piezoelectric material whose tip
  mass is subjected to some vibration and is also electromechanically
  coupled with the current flowing in the electric circuit.}  Although
piezoelectric vibration harvesters are typically studied in sinusoidal
conditions, at frequencies within their resonance band~\cite{H1,H2},
{ as mentioned above it is also important to} consider
broadband vibrations, which can be modelled as white
noise~\cite{halv,A,C,D,noi1,noi2,noi3}. {From a more theoretical
  perspective,} stochastic driving makes this kind of system an
interesting instance where results from {the general
  theory of} stochastic thermodynamics~\cite{seifertrev} can be
applied, as, for instance, discussed in~\cite{noi1}. {This
  theory represents an attempt to generalize the concepts of standard
  thermodynamics, such energy, heat, and entropy, to systems where
  fluctuations play a central role and cannot be neglected. For
  instance, in~\cite{noi1}, it is shown how a fluctuating power can be
  defined according to the prescriptions of stochastic thermodynamics
  for this kind of system. Moreover, as detailed below in the model equations,} the coupling
between tip mass velocity and electrical current introduces a feedback
mechanism in the system {that can be represented as memory effects}~\cite{loos2021stochastic}.

We will present results from experiments {performed on the
  piezoelectric harvesters} with two different setups {for
  the load electrical circuit}: (I) a linear configuration, which
allows for analytical computations, as previously studied
in~\cite{noi1}, in order to fix the main relevant parameters of the
model; (II) a nonlinear configuration, featuring a diode bridge as the
output circuit.  We present stochastic models that can capture the
essential physical processes and mechanisms underlying the energy
conversion in these configurations, taking into account various
parameters, such as the mechanical properties of the piezoelectric
harvester and the electrical characteristics of the harvesting
circuit. These models are mainly based on an underdamped Langevin
equation, which describes the dynamics of the tip mass of the
piezoelectric material, subjected to the viscous drag of the air and
confined by an elastic potential. The tip mass velocity is also
electromechanically coupled with the current flowing in the electric
circuit, as described in the following in detail. Moreover, the
piezoelectric material is connected with a shaker, which represents the
source of vibrations which are converted to electrical power.  In
particular, regarding the nonlinear configuration, we will show that a
simplified effective model for the diode bridge rectifier is
sufficient to reproduce the average extracted power.

Moreover, the system under study allows us to address an interesting
issue related to the inference of the non-equilibrium properties of
the system from a temporal series of
data~\cite{PhysRevResearch.4.043103}. We will show that, at variance
with the linear case discussed in~\cite{noi3}, in the nonlinear configuration the
time asymmetry can be revealed by the analysis of high-order
correlation functions of a single variable.

This paper is organised as follows. In Section~\ref{sec:exp}, we
describe the experimental setting, and provide details on the two
configurations considered. In Section~\ref{sec:theory}, we present the
stochastic theoretical models, based on the Langevin equation. In
particular, we propose an effective equation for the modelling of the
diode bridge rectifier. We compare analytical and numerical results
with the experimental data, finding very good agreement in a wide
range of parameters. In Section~\ref{sec:noneq}, we discuss the problem
of inferring the time asymmetry of the system from data
analysis. Finally, in Section~\ref{sec:conc} we present some
conclusions and perspectives for future works.

\section{Experimental Setup}
\label{sec:exp}

\begin{figure}[h]
  
      \includegraphics[width=0.45\textwidth]{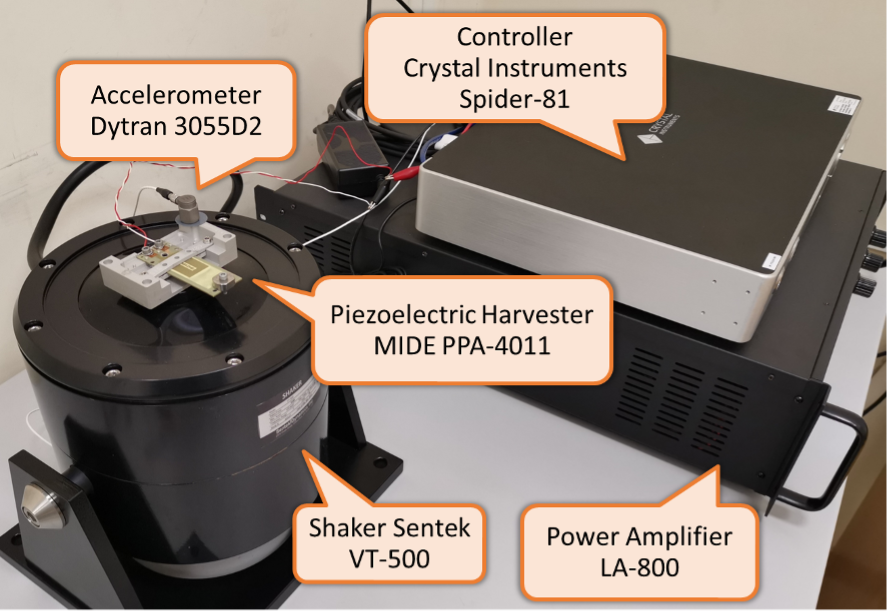}
      \caption{Picture of the experimental setup.}
   \label{fig_exp1}
 \end{figure}

The employed experimental setup is shown in the photo reported in Figure \ref{fig_exp1}. In particular, the considered harvester is the commercial piezoelectric device MIDE PPA-4011 (by MIDE Technology, Woburn, MA, USA) 
. The harvester was mounted in a cantilever resonant structure and placed on a shaker, the VT-500 by Sentek (by Sentek Dynamics Santa Clara, CA, USA), which was used as the source of the desired vibrations. The shaker driving current was provided by a Power Amplifier LA-800 (by Sentek Dynamics Santa Clara, CA, USA) whose control signal was generated by closed-loop vibration control implemented by a Crystal Instruments Spider-81 (by Crystal Instruments Santa Clara, CA, USA) measuring the shaker acceleration by means of an accelerometer Dytran 3055D2 (by Dytran (HBK) Chatsworth, CA, USA). 

\begin{figure}[h]
   \center
   \includegraphics[width=1.\textwidth]{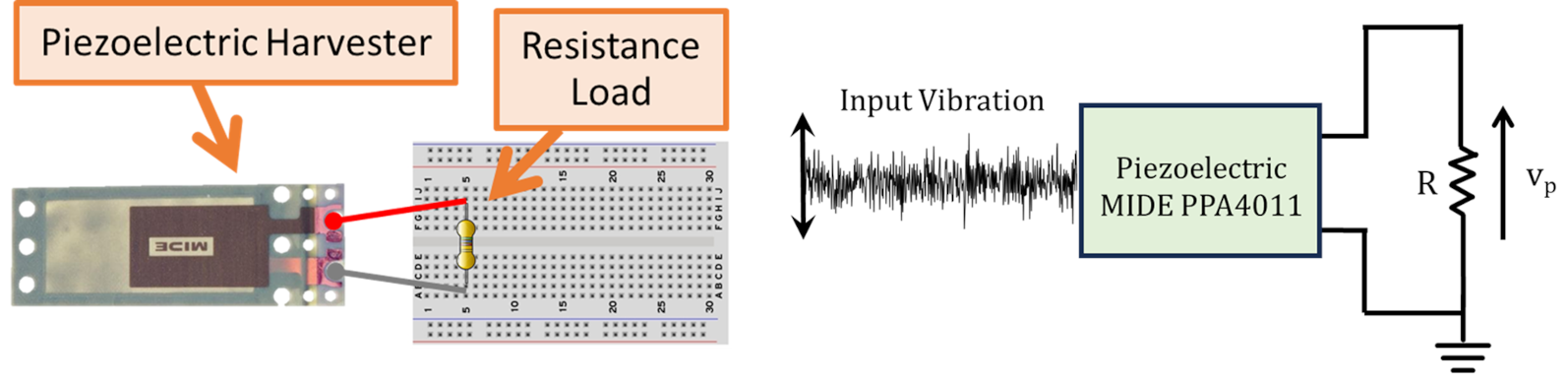}
      \caption{Schematic representation
 ({\bf left}) and circuital representation ({\bf right}) of the piezoelectric harvester loaded by the resistive load.}
   \label{fig_exp23}
 \end{figure}

\begin{figure}[h]
   \center
   \includegraphics[width=1.\textwidth]{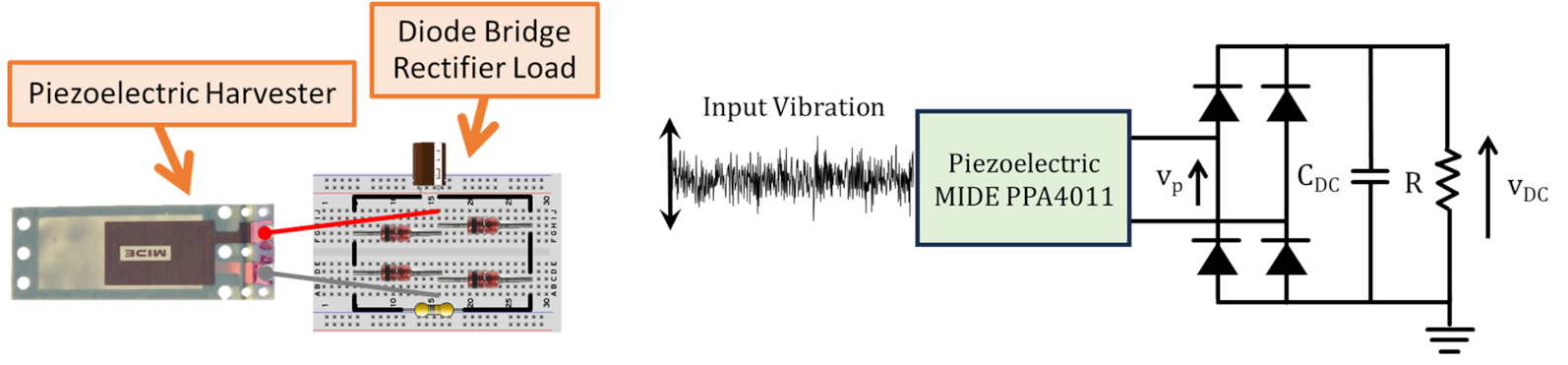}
      \caption{Schematic representation
 ({\bf left}) and circuital representation ({\bf right}) of the piezoelectric harvester loaded by the diode bridge rectifier load.}
   \label{fig_exp45}
 \end{figure}

Schematic and circuital representations of the experimental setup
showing the different considered harvester loads are reported in
Figures \ref{fig_exp23} and \ref{fig_exp45}. In particular, the
piezoelectric harvester was forced by broadband vibrations of Gaussian
type, with sampling rate $f_s = 5$ kHz and different standard
deviations (0.8 g and 1 g, where g is gravity acceleration). As
shown in Figure \ref{fig_exp23}, firstly a linear resistive load
was considered and the voltage $v_p$ across the load resistance $R$
was measured and recorded for different values of $R$. The second
considered harvester load was a nonlinear load consisting of a diode
bridge rectifier. Such a kind of circuit is typically employed in
piezoelectric vibration energy harvesting applications, in cases with
both laboratory prototypes and commercial devices
\cite{brenes2020maximum,dicken2012power,data1,data2,data3}, with the
aim of carrying out the AC/DC conversion, which is necessary for
supplying electronic DC loads (like sensors of a wireless sensor
network). As shown in Figure \ref{fig_exp45}, the diode bridge
rectifier (made of four 1N4148 diodes) is connected to the parallel
between a capacitor ($C_{DC} = 100~\mu$F) and a resistor with
resistance $R$. In this case, the voltage $v_p$ at the input of the
diode bridge rectifier and the voltage $v_{DC}$ across the load
resistance $R$ were measured and recorded for different values of $R$.

\section{Theoretical Model}
\label{sec:theory}

To describe the experimental setup, we consider
 the following stochastic model:
\begin{eqnarray}
  \dot{x}&=&v, \label{lang1}\\
  M\dot{v}&=&-K_sx -\gamma v -\theta v_p + M \xi, \label{lang2}\\
  C_p\dot{v}_p&=&\theta v - i_p, \label{lang3} \\
  i_p&=&f(v_p),
\end{eqnarray}
where $\xi$ is white noise with zero mean and correlation $\langle
\xi(t)\xi(t')\rangle = 2D_0\delta(t-t')$.  In the above equations, $x$
represents the displacement of the tip mass $M$, $v$ represents its velocity,
$\gamma$ represents the viscous friction due to air, $K_s$ represents the stiffness of the
cantilever in the elastic approximation, $v_p$ represents the voltage across the
load resistance $R$, $C_p$ represents the effective capacitance in the circuit,
$\theta$ represents the electromechanical coupling factor of the transducer,
$i_p$ represents the current flowing in the electrical circuit, and $f(v_p)$ represents the
characteristic current-voltage of the electrical load connected with
the piezoelectric harvester. Thermal fluctuations on the tip mass are
too small to affect its motion and are neglected. 

The explicit forms of the function $f(v_p)$ corresponding to the two considered experimental configurations are
\begin{equation}
f(v_p)=\frac{v_p}{R},
\end{equation}
for the linear case, which we denote by Configuration (I), and
\begin{eqnarray}
 f(v_p)&=&(I_{k0}+G v_{DC})\left(e^{\frac{v_p-v_{DC}}{2\eta V_T}}-e^{\frac{-v_p-v_{DC}}{2\eta V_T}} \right), \nonumber \\
  C_{DC}\frac{dv_{DC}}{dt}&=&-\frac{v_{DC}}{R}+(I_{k0}+G v_{DC})\left(e^{\frac{v_p-v_{DC}}{2\eta V_T}}+e^{\frac{-v_p-v_{DC}}{2\eta V_T}}-2 \right), \label{diode}
\end{eqnarray}
for the nonlinear (diode bridge rectifier) case, denoted as Configuration (II). Here, $\eta=1.94$ for the considered diodes and $V_T=25$ mV is the thermal voltage.

\subsection{Linear Load}

Case (I) has been carefully studied in previous works~\cite{noi1,noi2,noi3} and, due
to the linear nature of the model, allows for analytical treatment. It
is considered here in order to fit some of the model parameters, which
are then kept fixed in setup (II). We report the explicit
expression for the average output power extracted by the harvester,
which corresponds to the heat dissipated into the load resistance $R$
per unit time
\begin{equation}\label{pharv0}
P^{(I)}_{harv}=
\frac{1}{R}\langle v_p^2\rangle=\frac{D_0 M^2 R \theta^2}{ M (\gamma + R \theta^2) + 
  C_p R \gamma (C_p K_s R + \gamma + R \theta^2)},
\end{equation}
where the symbol $\langle\cdots\rangle$ denotes an average over noise
in the stationary state.  Details on the analytical solution of 
model (I) can be found in~\cite{noi1}.

\begin{figure}[h]

\centering %% If there is a figure in wide page, please release command \centering

%\centering
  \includegraphics[scale=0.32, clip=true]{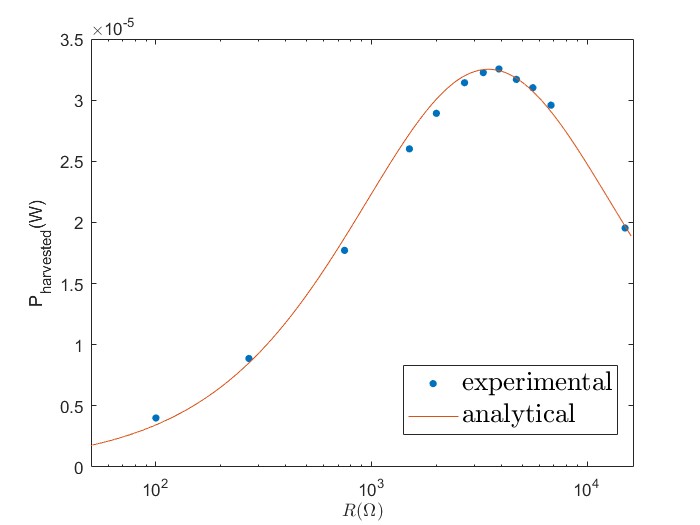}
  \includegraphics[scale=0.32, clip=true]{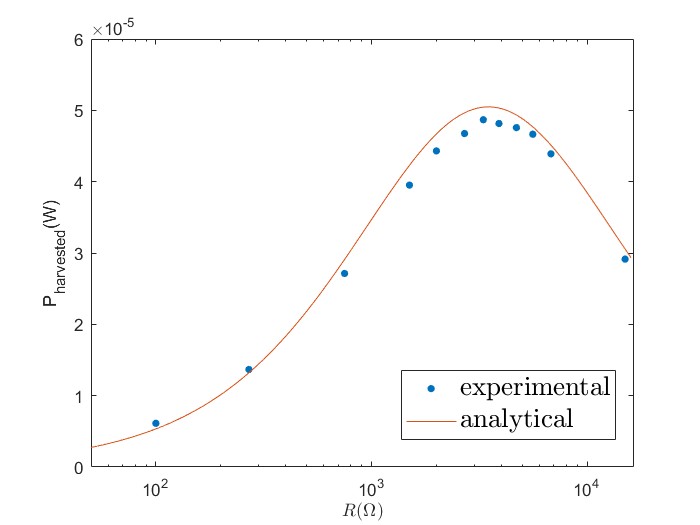}

  \caption{{Comparison} between the extracted power $P^{(I)}_{harv}$ in the linear setup (I) measured in experiments (dots) and the analytical prediction of Equation~(\ref{pharv0}) (line),
  with the values of the parameters reported in the text. ({\bf Left}): acceleration $a=0.8$ g. ({\bf Right}): acceleration $a=1.0$ g. The parameters are fitted to the case where $a=0.8$ g.}
\label{Fig:1}
\end{figure}

The noise amplitude $D_0$ is related to the shaker acceleration $a$
and to the sampling rate $1/\Delta t$ of the input signal,
$D_0=a^2\Delta t/2$, where $\Delta t=1/f_s=0.0002$ s. The ratio of
parameters $K_s/M$ is fixed by the characteristic frequency of the
device, with is $\sqrt{K_s/M}=2\pi\times 100$ Hz. The capacitor $C_p\sim 410$ nF is measured by using an LCR meter U1733C by Keysight Technologies, Colorado Springs, CO, USA.
  The other parameters are
fitted to the experimental data exploiting the analytical
expression~(\ref{pharv0}) as a function of the load resistance
$R$. For the case where $a=0.8\times 9.81~{\text{m/s}}^2$, we obtain the following
values: $M=0.0112\pm 0.0005$ Kg, $\theta=0.0172 \pm 0.0005$ N/V, and $\gamma=0.660 \pm 0.005$ Kg/s. This
set of parameters is used also for other values of the shaker
accelerations used in the experiments, $a=1.0\times 9.81 ~{\text{m/s}}^2$. In
Figure~\ref{Fig:1}, we report the experimental data and the analytical
curve for case (I). We observe a non-monotonic behaviour of the
extracted power as a function of the load resistance, with a peak
corresponding to the optimal load $R^*\sim {\text{3000--4000}}~ \Omega$. As
predicted by analytical Formula~(\ref{pharv0}), this value is
independent of the forcing acceleration, as shown in the right panel
of Figure~\ref{Fig:1}. In this figure, we also observe that the
parameter values reported above (fitted to the case where $a=0.8$ g) describe quite well the behaviour for a different value of the
acceleration ($a=1.0$ g).

%%%%%

\begin{figure}[h]

\centering 
  \includegraphics[scale=0.32, clip=true]{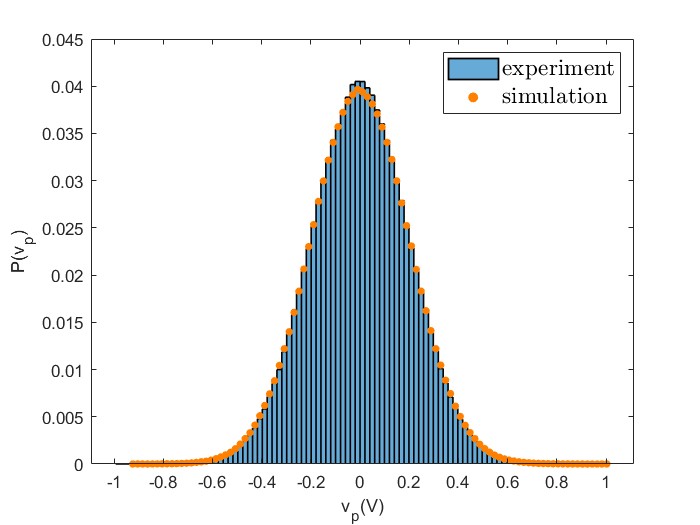}
  \includegraphics[scale=0.32, clip=true]{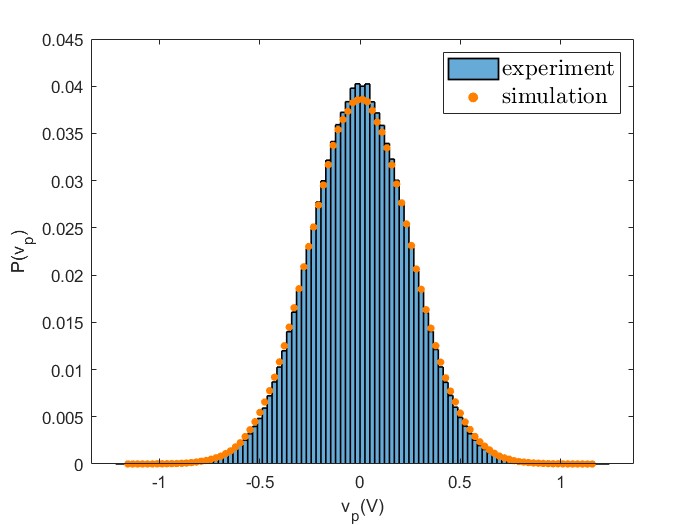}

\caption{Probability
 distributions of the voltage $v_p$ measured in
  experiments (histogram) and in numerical simulations (dots) in the
  case of load resistance $R=1500~ \Omega$, for Configuration (I).  ({\bf Left}): acceleration $a=0.8$ g. ({\bf Right}): acceleration $a=1.0$ g. Other values of the resistance
  show similar behaviours.}
\label{Fig:2}
\end{figure}

As already shown in previous works~\cite{noi1}, the efficacy of the
linear model in describing the real system is not limited to mean
values of the relevant quantities, but also extends to the
fluctuations in the voltage $v_p$, as reported in
Figure~\ref{Fig:2}. Here, we show the histograms of the measured values
of $v_p$ in experiments and in numerical simulations for a fixed value
of the load resistance $R=1500~ \Omega$, for the two values of the
considered acceleration, which are in very good agreement. As expected
from the linearity of the model, these distributions show a Gaussian
shape. The analytical expression for the variance of the Gaussian as a
function of the model parameters is reported in~\cite{noi1}. For other
values of the load resistance, we find similar behaviours.

%%%%%

\subsection{Nonlinear Load}

An analytical treatment is not possible for the model of Configuration (II), due to the
strong nonlinearities appearing in the equations.
Therefore, here we focus on numerical simulations in order to
investigate the accuracy of the proposed modelling, in particular for
what concerns the description of the diode bridge.
Here, we simplify the treatment, introducing two effective parameters,
$I_{k0}$ and $G$, that appear in Equation~(\ref{diode}). The effective values of these parameters used in numerical simulations are $I_{k0}=9\times 10^{-9}$~A and $G=-133\times 10^{-11}~\Omega^{-1}$. 

\begin{figure}[h]

\centering

  \includegraphics[scale=0.32, clip=true]{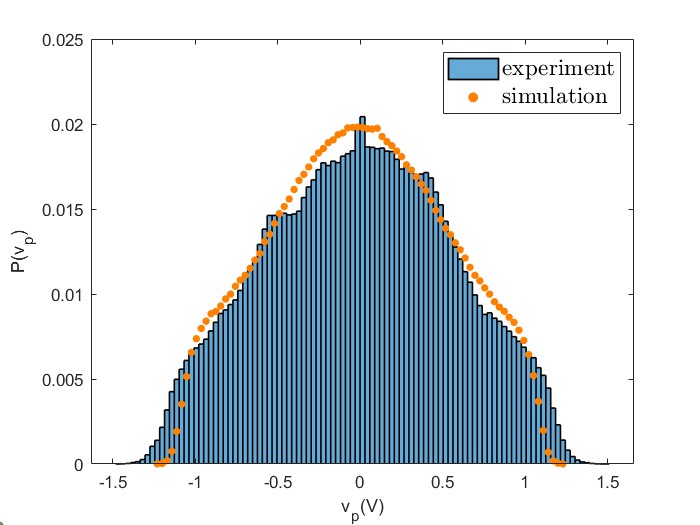}
  \includegraphics[scale=0.32, clip=true]{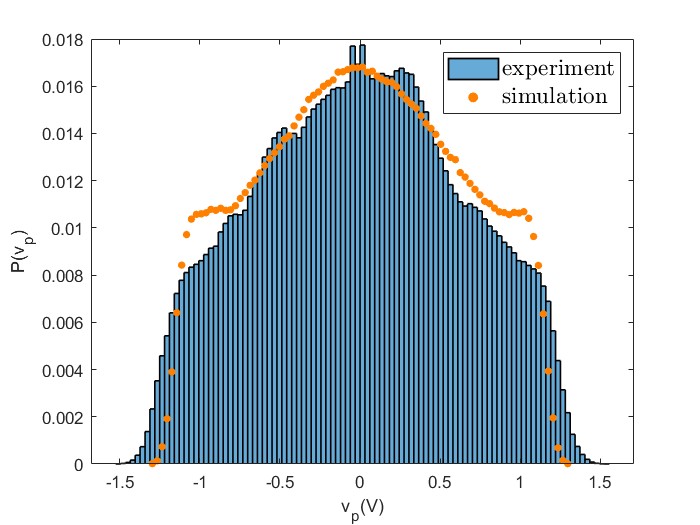}

\caption{{Probability} 
 distribution of the voltage $v_p$ measured in experiments and numerical simulations for $R=3300~ \Omega$ for $a=0.8$ g 
 ({\bf left}) and $a=1.0$ g ({\bf right}).}
\label{Fig:2a}
\end{figure}

\begin{figure}[h]

\centering
  \includegraphics[scale=0.32, clip=true]{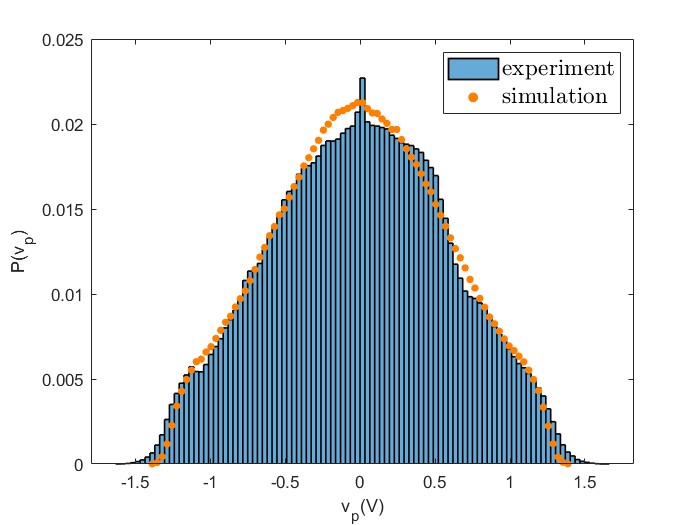}
  \includegraphics[scale=0.32, clip=true]{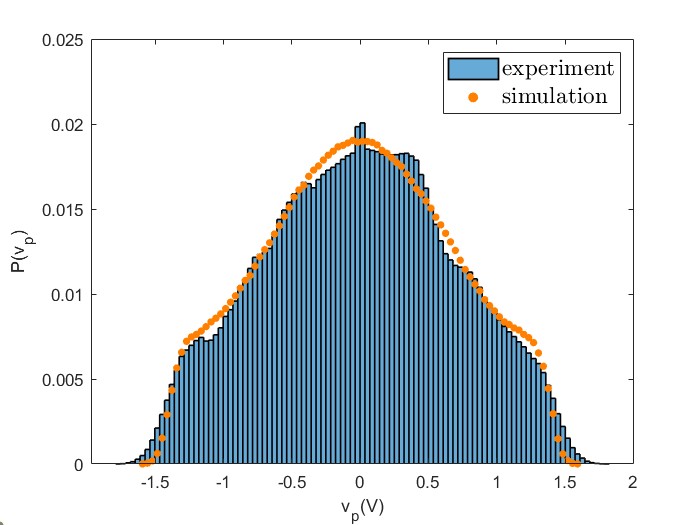}

\caption{{Probability} 
 distribution of the voltage $v_p$ measured in experiments and numerical simulations for $R=20,000 ~\Omega$ for $a=0.8$ g
 ({\bf left}) and $a=1.0$ g ({\bf right}).}
\label{Fig:2b}
\end{figure}

\begin{figure}[h]

\centering %% If there is a figure in wide page, please release command \centering

  \includegraphics[scale=0.32, clip=true]{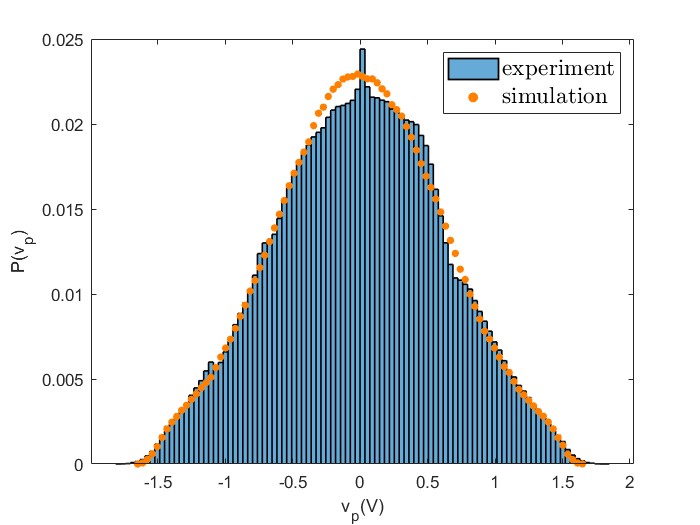}
  \includegraphics[scale=0.32, clip=true]{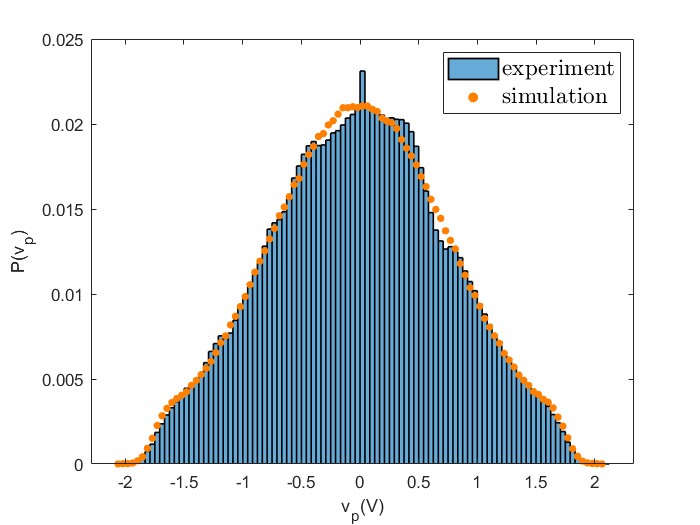}

\caption{{Probability} 
 distribution of the voltage $v_p$ measured in experiments and numerical simulations for $R=100,000~ \Omega$ for $a=0.8$ g ({\bf left}) and $a=1.0$ g ({\bf right}).}
\label{Fig:2c}
\end{figure}

As first, we consider the voltage $v_p$ and compare the probability
distributions measured in experiments with those obtained from
numerical simulations.  The good agreement between experiments and
simulations for different values of the load resistance $R$ observed
in Figures~\ref{Fig:2a}--\ref{Fig:2c} shows that fluctuations are well
described by the nonlinear model. In particular, we note that these
distributions have a shape very different from a Gaussian, due to the
nonlinearity of the system. These differences appear more pronounced
at small values of $R$ and are characterised by large tails. Similar
behaviours are observed for both the values of the considered
acceleration and for other values of the resistance $R$.

%%%%%

%%%%%

%%%%%

\begin{figure}[h]

\centering %% If there is a figure in wide page, please release command \centering

  \includegraphics[scale=0.32, clip=true]{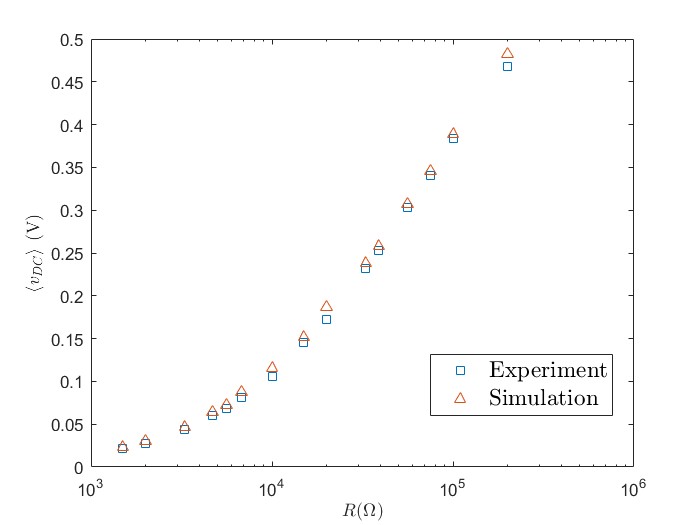}
  \includegraphics[scale=0.32, clip=true]{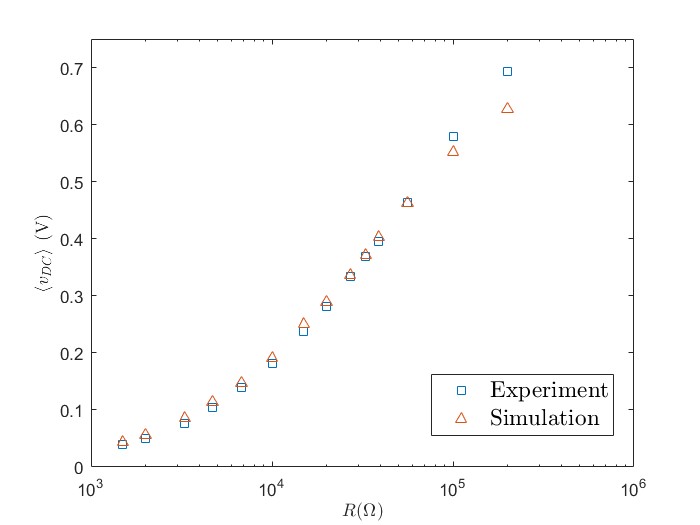}

\caption{Average $\langle v_{DC}\rangle$ as a function of $R$ for $a=0.8$ g ({\bf left}) and $a=1.0$ g ({\bf right}).}
\label{Fig:3}
\end{figure}

We then consider the quantity $v_{DC}$, which is relevant for the harvested power 
through the expression
\begin{equation}
P^{(II)}_{harv}=\frac{\langle v_{DC}^2\rangle}{R}.
\end{equation}
In order to test the accuracy of the model, we first consider the
average value $\langle v_{DC}\rangle$, which is reported in
Figure~\ref{Fig:3} for different values of the load resistance. Good
agreement is found between experiments and numerical simulations. We
observe an increasing behaviour with $R$. {This is
  expected since, the more $R$ increases, the closer the system
  approaches open-circuit operating conditions.}

%%%%%

\begin{figure}[h]

\centering %% If there is a figure in wide page, please release command \centering

  \includegraphics[scale=0.32, clip=true]{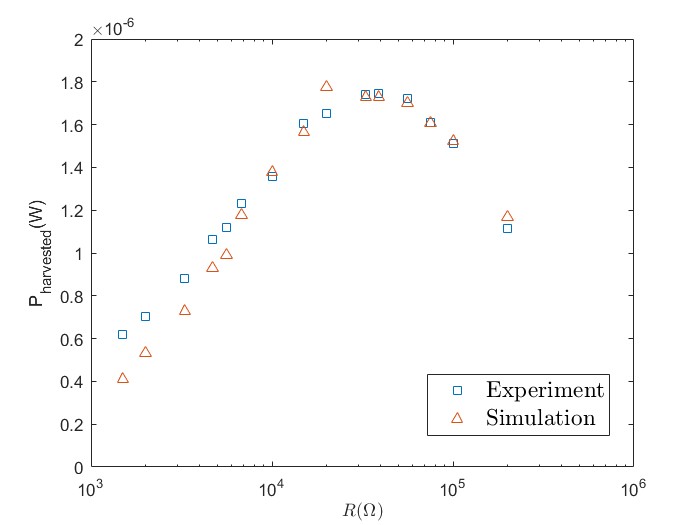}
  \includegraphics[scale=0.32, clip=true]{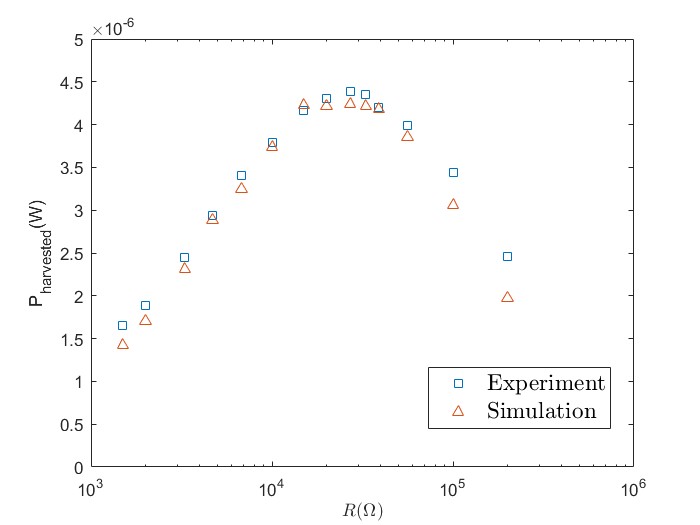}

\caption{{Extracted} 
 power $P^{(II)}_{harv}=\langle v_{DC}^2\rangle/R$ as a function of $R$ for $a=0.8$ g ({\bf left}) and $a=1.0$ g ({\bf right}).}
\label{Fig:4}
\end{figure}

We then consider the harvested power, which is reported in
Figure~\ref{Fig:4}, and which involves the fluctuations $\langle
v_{DC}^2\rangle$. {This is the most important quantity,
  because it represents the energy that can be practically used to power
  other devices.} Again, with the considered parameters, good
agreement is found, in the whole range of values of $R$, {
  between the numerical simulations and experimental results}. We observe
a non-monotonic behaviour characterised by a maximum for the optimal
load resistance $R^*\sim 30,000~ \Omega$. We note that the maximum
extracted power is smaller with respect to the linear
case. {However, as mentioned above, this kind of load
  circuit is necessary in practical applications to carry out the
  AC/DC conversion for supplying electronic loads.}

%%%%%

\begin{figure}[h]

\centering %% If there is a figure in wide page, please release command \centering

  \includegraphics[scale=0.32, clip=true]{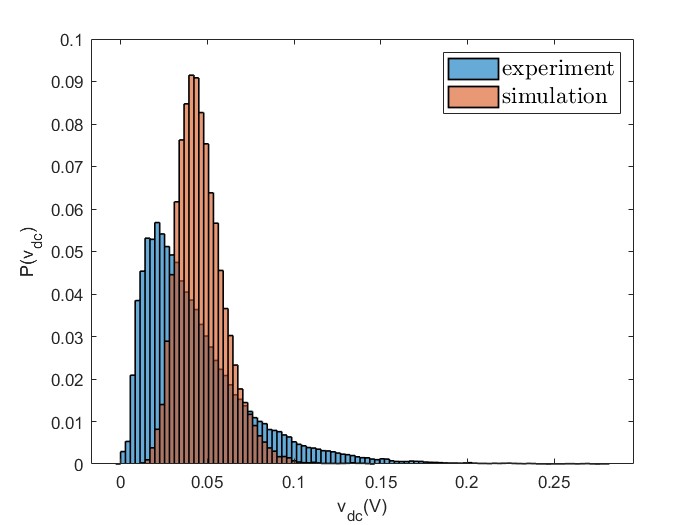}
  \includegraphics[scale=0.32, clip=true]{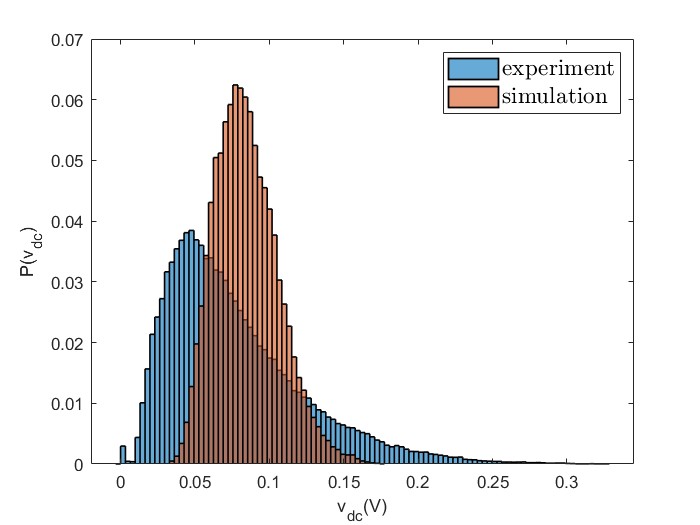}

\caption{Probability distributions  of the voltage $v_{DC}$ measured in experiments and numerical simulations for $R=3300~ \Omega$  for $a=0.8$ g ({\bf left}) and $a=1.0$ g ({\bf right}).}
\label{Fig:5a}
\end{figure}

Finally, in Figure~\ref{Fig:5a} we compare the probability distributions
for the quantity  %EE: Please check intended meaning has been retained Author: OK
$v_{DC}$ measured in experiments and in simulations
for $R=3300~ \Omega$. We observe that, even if average and variance are
well described by the model, the whole shape of the distributions is
not accurately reproduced. Other values of $R$ show similar
behaviours. This is probably due to the simplifications introduced in
the model to represent the diode bridge. {Our analysis,
  therefore, shows that the proposed model is a good compromise
  between accuracy and simplicity, allowing us to reproduce the
  behaviour of the main quantities, without the introduction of a large
  number of effective parameters.}

%%%%%

\subsection{Time-Correlation Functions and Temporal Asymmetry}
\label{sec:noneq}

The system considered in our study allows us to also address a
theoretically interesting problem, consisting of the assessment of the
non-equilibrium nature of the system from partial information, namely
from the measurement of only some degrees of freedom. This is a
non-trivial issue, because in experiments one usually cannot directly
access the dynamics of all the relevant quantities and the evaluation
of the temporal (a-)symmetry of the system behaviour can be
difficult. The question plays a central role in the general problem of
finding a good model from data, where knowledge of the
equilibrium/non-equilibrium properties of the system can provide a
useful base for modelisation.  Several theoretical tools can be used
to assess such features, such as entropy production
measurements~\cite{manikandan2020inferring}, violations of the
fluctuation--dissipation relations~\cite{puglisi2017temperature}, and high-order correlation functions~\cite{pomeau1982symetrie}.  In particular,
as recently discussed in~\cite{PhysRevResearch.4.043103}, the Gaussian
nature of the model can hide the non-equilibrium features of the
dynamics when only one degree of freedom is considered, leading to the
necessity of the analysis of cross-correlation between two variables
to assess the non-equilibrium behaviours. This point was considered in
detail in a previous paper of some of the present authors~\cite{noi3},
where the linear model of Configuration (I) was considered, and indeed
it was shown that the single measurement of the voltage $v_p$ was not
enough to unveil the non-equilibrium system dynamics. In that case,
the cross-correlation of $v_p$ with another variable related to the
displacement of the mass tip of the piezoelectric was considered.

\begin{figure}[h]

\centering %% If there is a figure in wide page, please release command \centering

  \includegraphics[scale=0.32, clip=true]{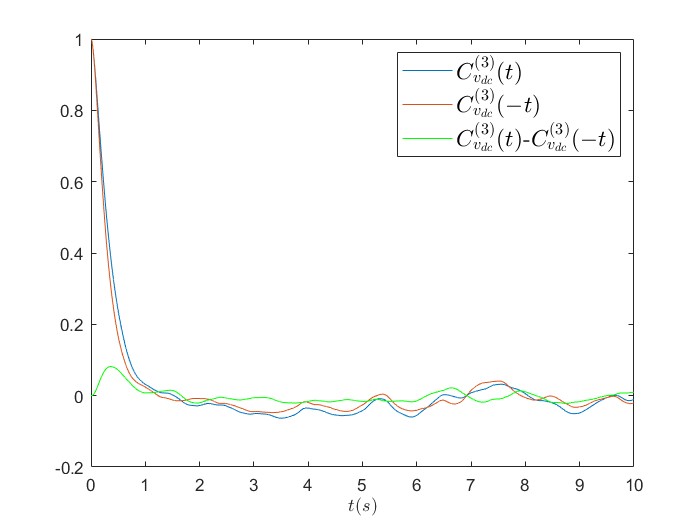}
  \includegraphics[scale=0.32, clip=true]{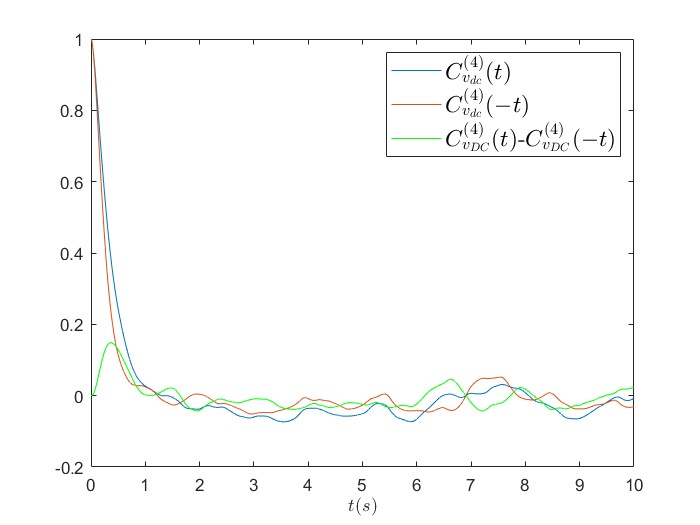}

\caption{{Multi-point} 
 correlation functions of the voltage $v_{DC}$, $C_{v_{DC}}^{(3)}(t)$ (left panel) and $C_{v_{DC}}^{(4)}(t)$  measured in experiments with $R=3300 ~\Omega$ for $a=0.8 ~g$. The time asymmetry is revealed by the difference with respect to the same functions computed by inverting the time argument.}
\label{Fig:6}
\end{figure}

Here, exploiting the nonlinear components of the system in
Configuration (II), we show that from the analysis of the time series
of a single variable it is possible to reveal the time asymmetry of
the dynamics. In particular, we define the connected three- and four-point correlation functions of the voltage $v_{DC}$ as follows:
\begin{equation}
  C_{v_{DC}}^{(3)}(t)=\frac{\langle v_{DC}(t) v_{DC}(0)^2\rangle-\langle v_{DC}\rangle \langle v_{DC}^2\rangle}{\langle v_{DC}^3\rangle-\langle v_{DC}\rangle \langle v_{DC}^2\rangle},
\end{equation}

\begin{equation}
  C_{v_{DC}}^{(4)}(t)=\frac{\langle v_{DC}(t) v_{DC}(0)^3\rangle-\langle v_{DC}\rangle \langle v_{DC}^3\rangle}{\langle v_{DC}^4\rangle-\langle v_{DC}\rangle \langle v_{DC}^3\rangle}.
\end{equation}
The choice of these kinds of correlation functions is dictated by the
observation that simple two-point autocorrelation is always
time-symmetric by definition and, therefore, higher-order functions have
to be taken into account. In Figure~\ref{Fig:6}, we report
$C_{v_{DC}}^{(3)}(t)$ (left panel) and $C_{v_{DC}}^{(4)}(t)$ (right
panel) measured in experiments with $R=3300 ~\Omega$ for $a=0.8~ g$.
Other values of parameters show similar behaviours. We clearly see the
time asymmetry of the dynamics from the difference between these
functions and those obtained inverting the time argument.

%%%%%

\begin{figure}[h]
%\centering

\centering %% If there is a figure in wide page, please release command \centering

  \includegraphics[scale=0.32, clip=true]{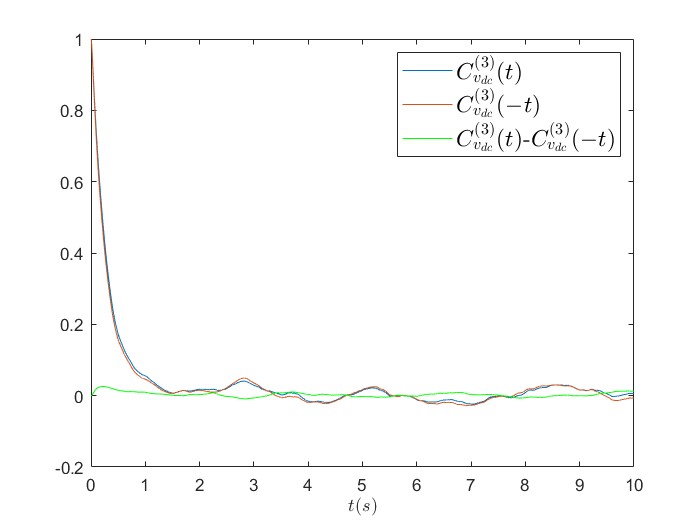}
  \includegraphics[scale=0.32, clip=true]{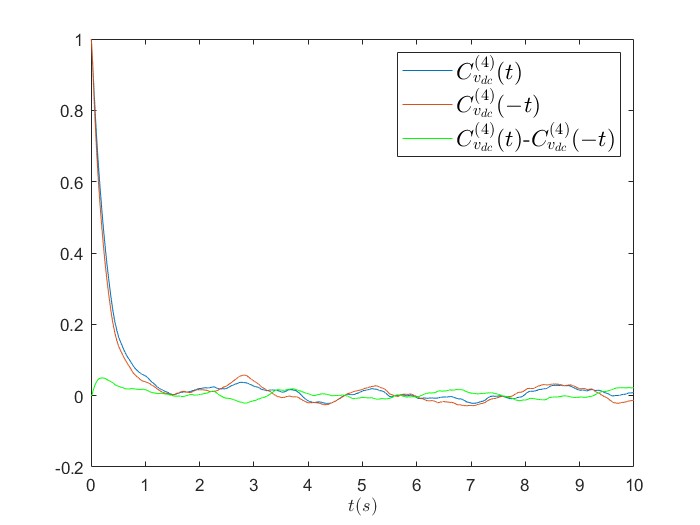}

\caption{{{Multi-point} 
 correlation functions of the voltage $v_{DC}$, $C_{v_{DC}}^{(3)}(t)$ (left panel) and $C_{v_{DC}}^{(4)}(t)$  measured in numerical simulations with $R=3300 ~\Omega$ for $a=0.8 ~g$. The time asymmetry is revealed by the difference with respect to the same functions computed by inverting the time argument.}}
\label{Fig:7}
\end{figure}

{ In order to verify in more detail how much the model can
  capture the non-equilibrium properties of the real system, we also
  compute the same high-order correlation functions from numerical
  simulations, for the same parameters used before. In
  Figure~\ref{Fig:7}, we report $C_{v_{DC}}^{(3)}(t)$ (left panel) and
  $C_{v_{DC}}^{(4)}(t)$ (right panel), which show a qualitative
  behaviour similar to what we observed in the case of experimental data:
  the presence of a peak at small times, in particular, even if less
  evident with respect to the experimental data, signals the temporal
  asymmetry of the dynamics, and, therefore, confirms the
  non-equilibrium nature of the theoretical model. The quantitative
  disagreement is probably due to the simplifications introduced in
  the modelling of the diode bridge, as already discussed before.}

%%%%%

\section{Conclusions}
\label{sec:conc}

{This work focuses on the modelisation} of a piezoelectric
energy harvester driven by random broadband vibrations.{
  We analyse data obtained} in two different experimental setups: a
linear configuration, characterised by a single load resistance in the
output circuit, and a nonlinear one, where a diode bridge is
considered. {The theoretical model proposed to describe
  the experimental results relies on} a stochastic equation, based on
an underdamped Langevin equation for the tip mass dynamics
{of the piezoelectric material}, electromechanically
coupled with the output electrical circuit. As also shown in previous
studies~\cite{noi1,noi2,noi3}, the system in the linear setup is very
well fitted by the theoretical model. Here, we consider this
configuration to fix some of the model parameters. We then analyse the
accuracy of the model when the diode bridge is studied.{
  This nonlinear configuration is more realistic for practical
  implementations but cannot be treated analytically. Therefore, we
  perform extensive numerical simulations of the model to investigate
  the role of the parameters, in particular on the dependence of the
  extracted power on the load resistance}, finding a good agreement
between experiments and numerical simulations. {In
  particular, our approach} proposes a simplified model for the
description of the diode circuit, which allows us to recover the
experimental behaviour with a small number of
parameters. {The main result of our analysis consists of
  the observation of} a non-monotonic behaviour of the extracted power
as a function of the load resistance, identifying the optimal value at
which the power is maximised.

We also address the {more theoretical} issue related to
the {characterisation} of non-equilibrium features in the
system from the analysis of a time series of a single observable variable. 
At
variance with previous studies performed in the linear setup, we show
here that, as expected from general arguments, in the case of
nonlinear systems, to demonstrate a temporal asymmetry of the dynamics,
it is not necessary to measure cross-correlations between two
different variables, but the computation of high-order correlation
functions of a single variable is sufficient.

Our study extends the validity of the proposed stochastic model to
treat piezoelectric energy harvesters driven by broadband vibrations
to nonlinear setups and represents a physical example where the
properties of non-equilibrium fluctuations {in systems
  with feedback dynamics} can be studied.

%%%%%%%%%%%%%%%%%%%%%%%%%%%%%%%%%%%%%%%%%%
\vspace{6pt} 

%%%%%%%%%%%%%%%%%%%%%%%%%%%%%%%%%%%%%%%%%%
%% optional
%\supplementary{The following are available online at \linksupplementary{s1}, Figure S1: title, Table S1: title, Video S1: title.}


%%%%%%%%%%%%%%%%%%%%%%%%%%%%%%%%%%%%%%%%%%
\acknowledgments{This research
 was supported in part by Unione Europea - Next Generation EU and MUR in the framework
of PRIN 2022 under grant 20222RWCJJ (HEAVEN), in part by Unione Europea - Next Generation EU and MUR
in the framework of PRIN 2022 under grant 2022Z8C472 (AMPERE) and in part by Unione Europea - Next
Generation EU and MUR in the framework of PRIN 2022 PNRR under grant P202244448 (ESPERI).}

%%%%%%%%%%%%%%%%%%%%%%%%%%%%%%%%%%%%%%%%%%

%%%%%%%%%%%%%%%%%%%%%%%%%%%%%%%%%%%%%%%%%%

%%%%%%%%%%%%%%%%%%%%%%%%%%%%%%%%%%%%%%%%%%
\end{document}